# Numerical modelling of MgB$_2$ conductors for high power AC transmission

F. Grilli[a*], A. Chervyakov[b], V. Zermeno[a], A. Marian[b], G. Grasso[c], W. Goldacker[a], C. Rubbia[b]

[a]*Karlsruhe Institute of Technology, Karlsruhe, Germany*
[b]*Institute for Advanced Sustainability Studies, Potsdam, Germany*
[c]*Columbus Superconductors, Genova, Italy*

**Abstract**

Cables made of MgB$_2$ superconductors are currently explored as a viable solution for transporting high electrical power in the AC regime. In order to be competitive against the DC solution, the cables need to have an acceptable level of AC losses. In this contribution, we discuss the main aspects relevant for designing a cable with a sufficiently low AC loss level. To this end, we perform finite-element-method (FEM) simulations to determine the current and field distributions and calculate the AC losses of such cable configuration. For current capacities of 2-5 kA (peak), power cables are assembled from a relatively small number of MgB$_2$ strands. The performance of such cables strongly depends on the current and field distributions, which are in turn influenced by the number and the arrangement of the superconducting components and also by the magnetic properties of supporting materials. Numerical simulations can help to test different cable configurations and provide important insights for optimizing the cable's design. The numerical model includes the field dependence of the superconductor's critical current density $J_c(B)$ as well as the non-linear properties of magnetic materials.

## 1. Introduction

Since its discovery in 2001 [1], the $MgB_2$ superconductor has become an attractive candidate for various power applications at temperatures of liquid hydrogen or gaseous helium [2]-[4]. Apart from good transport properties, it demonstrates the same efficiency as already existing high-temperature superconductors (HTS) at a lower cost, with better mechanical stability, and more significant compactness in the structure [2],[4]. It also surpasses conventional low-temperature superconductors (LTS) by a higher temperature margin and an increased stability, two characteristics that are especially important for the transmission of electrical energy of a very high power [2],[4].

Recently, the $MgB_2$ superconductor has been proposed for transferring and balancing many Gigawatts of electrical power over long distances [5]. The conceptual designs of the DC power transmission system developed in [5] aim at transporting very large power with no electrical resistivity at relatively moderate voltages and higher currents when compared to conventional transmission lines.

In pure DC conditions, the losses of superconducting lines are mostly associated with the AC/DC conversion and the cooling, whereas the AC loss is only due to ripples. It becomes however a major problem for superconducting transmission of AC electrical power, for instance, after insertion of superconducting links into a power grid with no AC-DC-AC conversion. In order to become competitive, the design of any superconducting power cable for AC transmission should be carefully optimized to provide a minimal AC loss.

In this paper, we use the recent advancements in $MgB_2$ superconductors with the aim of designing a power cable with a minimal level of AC losses for future AC power applications. The numerical modelling of current and field distributions provides the base for initial assessment of the optimum cable design. Our results show that a properly designed MgB$_2$ power cable is a feasible solution for transportation of electrical power in the AC regime.

## 2. Cable geometry, numerical model and AC loss results

For an initial assessment, we design a 1.8 kA power cable (2.5 kA peak) for transmission of alternating current in self-field conditions. The cable is assembled from twisted MgB$_2$ wires. The reference wire of 1.1 mm diameter comprises the recent developments of the MgB$_2$ wires with uniform properties over long lengths made by Columbus Superconductors Spa (Italy) (see inset of Figure 1). It consists of a copper core, iron barrier and 12 twisted filaments immersed in a Monel matrix. The superconducting volume fraction of the wire is 0.15 and the critical current is 387 A at 20 K. This value was estimated in self-field conditions using the numerical method described in [6]. Typically, the total AC loss in a single wire is the sum of various loss components coming from the superconducting filaments, the matrix and the stabilizer. Understanding these contributions provides an important insight of the total AC loss. The total AC loss however should be calculated in the context of the cable design, rather than separately, since each strand inside the cable represents an electromagnetically interacting superconducting path.

The optimum cable design with an acceptable level of the AC loss value is achieved by establishing and accounting for the major factors affecting the total AC loss. The final design of the cable of 6.8 mm diameter is shown in Figure 2. This was obtained by simulating the current and field distributions in the cable's interior.

It was found that in the case when the filaments are coupled, the current is not evenly distributed among them (see Figure 3). Instead, the outermost filaments are saturated first, leaving the inner ones free of current. This resembles the current distribution in a large solid superconductor. On the other hand, decoupling of the filaments allowed a more homogeneous current distribution among them (see Figure 4). Decoupling also yielded lower AC losses estimates. To decouple the filaments, the $MgB_2$ strands in the cable should be twisted together. Twisting makes them electromagnetically equal and provides a better mechanical integrity of the cable.

In order to prevent flux penetration inside the central copper core (which causes ohmic dissipation), a protective layer of iron between the $MgB_2$ strands and the copper core is used as a magnetic screen. Finally, due to the relatively high generated self-field, the cable's critical current becomes lower than expected. Thus, to achieve a critical current capacity of about 4 kA that ensures an operating current of 2.5 kA (peak), one needs a sufficiently large number of wires, which we estimated to be 16.

The model for AC loss computation is based on the *H*-formulation of Maxwell equations solved by finite-element method (FEM) [7], aptly modified to account for the non-linear magnetic properties of some of the constituent materials [8]. The resistivity of the superconductor is modeled by means of a non-linear *E-J* characteristic

$$\rho(J) = \frac{E_c}{J_c(B)} \left|\frac{J}{J_c(B)}\right|^{n-1}$$

where $E_c=10^{-4}$ V/m, n=40 and $J_c(B)$ is the critical current density obtained from the experimental characterization of $MgB_2$ wires at 20 K for fields up to 1.7 T [9]:

$$J_c(B) = J_{c0}\left(1 - \frac{B}{B_1}\right)\left(1 + \frac{B}{B_0}\right)^{-\alpha}$$

with $B_0$=0.0339 T, $B_1$=2.91 T, and $\alpha$ =0.232. The value of $J_{c0}$ has been set to $3\cdot10^9$ A/m$^2$ to match the current densities of the best quality wires available at present. The electrical resistivities of other constituent materials are taken at 20 K as $\rho_{Cu}$=0.06 n$\Omega$.m, $\rho_{Fe}$=3.92 n$\Omega$.m, $\rho_{Monel}$=365· n$\Omega$.m. For iron and Monel, non-linear $\mu_r(H)$ functions were used in the calculation (Figure 1). The available data (see textbook [10] and references therein) were uploaded and interpolated by the software Comsol Multiphysics, where the FEM model was implemented.

The cable has a cylindrical geometry. Due to azimuthal symmetry, its cross-sectional area is split into 16 sectors, each including one strand. Only one of these sectors is modeled. This simplifies the simulation provided that periodicity conditions are applied on the periodical domain boundaries. The desired total transport current is injected into the whole conducting cross-sectional area (superconductor and metal parts) by means of a current constraint [7] and is let free to distribute in the different sub-domains. The twist of MgB$_2$ filaments is accounted for by applying the further constraint that the same current flows in each filament. It has to be noted that the value of this current is unknown, contrary to that of the total transport current. In this way, the simulated straight filaments represent the different positions assumed by a given twisted filament along one twist pitch in the real cable, although the simulations, being 2-D, cannot account for the axial component of the magnetic field. Without the use of these current constraints, the current tends to flow in the outer part of the cross-sectional area of the cable, using mostly the outer filaments and rapidly saturating them (with a consequent high energy dissipation), whereas the inner filaments are scarcely used, or even not utilized at all (Figure 3). However, it is interesting to note that also when all the filaments carry the same current (Figure 4), the current density distribution at their interior varies considerably: compare, for example, the leftmost and rightmost filaments in Figure 2 (number 1 and 7, according to the numbering scheme in the figure). In filament number 1, there is only a thin annulus with negative current density (the picture refers to $\omega t=3\pi/2$, i.e. at the negative peak of the total transport current). On the contrary, filament number 7, while carrying the same net current of number 1 in virtue of the current constraints applied as described above, has regions with both positive and negative current density (the latter being larger in order to match the imposed current constraint). This is a clear evidence of the presence of magnetization currents induced by the neighboring filaments and wires. This phenomenon is even more evident for the filaments number 4, 5, 9, and 10, whose cross-section is almost completely filled by the current. While the magnetization currents do not contribute to the net transported current, they do contribute to the losses. In fact, most of the energy dissipation occurring in the cable is due to them, as discussed later.

The magnetic field distribution in the cable is shown in Figure 5, and in the superconducting filaments in Figure 6. One can see from the pattern of the streamlines that the field generated by the transport current is mostly azimuthal, and is concentrated in the ferromagnetic parts, where it reaches the amplitude of about 600 mT for a transport current of 2.4 kA (peak). In the superconducting material, the maximum field is about 180 mT. Note the correlation between the magnetic field lines in the superconducting filaments and the patterns of the current density. The AC loss in the different filament positions is shown in Figure 7. As it can be expected from the current density and field distributions discussed above, the filaments with largest magnetization currents have the largest losses.

The calculated AC power loss as a function of the transport current is shown in Figure 8 for different choices of the parameter $J_{c0}$. For $J_{c0}=3\cdot10^9$ A/m$^2$, the AC loss is 2.2 W/m for a current of 2.4 kA (about 1.7 kArms), and 4 W/m for a current of 3.2 kA (about 2.3 kArms). The increase of the critical current capacity to $J_{c0}=10^{10}$ A/m$^2$ would reduce these values to 1.1 W/m and 2.6 W/m, respectively. At voltages of 220 kV AC, such a cable is capable of transferring 0.7 GVA of apparent power which amounts to 31 MVA at 10 kV AC. This is similar to efficiency of the 2.3 kA HTS power cable recently constructed for the Ampacity project [11-12]. It is also interesting to note that higher values of $J_{c0}$ do not necessarily correspond to lower AC losses. This is due to the fact that at high transport currents, the magnetic field generated by the neighboring strands completely saturates some (or even all) filaments. In that case, the currents in the filaments are essentially due to magnetization, as can be seen from Figure 9, where the data of Figure 8 are normalized by the square of the transport current. In this representation, the loss curve presents a peak, which can be put in relation to the full penetration of the magnetization currents into the superconducting filaments, similarly to the case of external magnetic field (see for example chapter 8 in [13]); once the superconductor is penetrated, the magnetization losses at a given field (i.e. at the field generated by a given current) increase with increasing $J_c$ [13]. The curve corresponding to $J_{c0}=10^{10}$ A/m$^2$ has the lowest losses because the critical current density is so high that the filaments are not saturated. Of course, having such a high value of $J_c$ is in principle desirable, but the cable would be operating at a very small fraction of its current-carrying capacity. On the other hand, operating with almost saturated filaments is not advisable because of the associated risk of instability. The determination of the optimal scenario in this respect requires additional investigations.

## 3. Conclusion and outlook

The analysis and simulations performed in this paper aimed at assessing the potential of $MgB_2$ cables for transporting high AC power: in particular, the goal was to estimate the AC loss level of a cable configuration consisting of round wires with state-of-the-art transport properties and already available in long lengths. For this purpose, a cable made of round multi-filamentary $MgB_2$ wires was analysed and simulated with a numerical model that includes the possibility of considering materials with non-linear magnetic properties and of controlling the current flowing in each conducting region. This latter feature makes it possible to have the same current flowing in each filament of an $MgB_2$ wire, a condition obtained in real cables by twisting the filaments.

It was found that in a cable composed from round wires, the AC loss is mostly due to magnetization currents induced by the azimuthal component of the magnetic field. For this reason, only a small fraction of the superconductor's cross-sectional area is utilized to transport a net current. The computed AC loss level is of the same order of magnitude as that of cables made of HTS tapes with a similar current capacity, although with a higher penalty due to the lower operating temperature. It has to be kept in mind, however, that other aspects such as material, manufacturing and maintenance costs as well as operating time of the cables should also be included in the comparison of the overall cost of the cable, and therefore the MgB2 solution would become competitive.

While improving the transport properties of the $MgB_2$ material is certainly useful for this purpose, careful studying all alternative configurations for the cable design in order to reduce the AC losses would be an important problem worth investigating in future works. In this work we neglected the hysteresis losses in the ferromagnetic materials (monel, iron): while these might have a non-negligible contribution to the losses of the cable, the low-cost of $MgB_2$ wires allows in principle the possibility of using more expensive non-ferromagnetic materials. This aspect can be the subject of future investigations as well.

**Acknowledgments**

This work has been partly supported by the Helmholtz-University Young Investigator Group grant VH-NG-617.

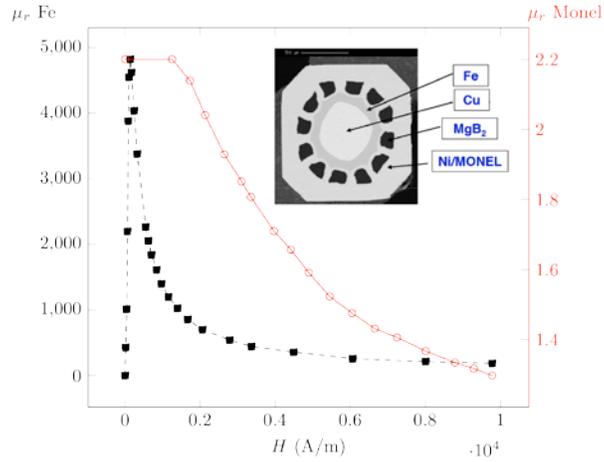

Figure 1. Figure Non-linear relative magnetic permeabilities of iron (full black squares) and Monel (open red circles) as a function of the applied field. The inset represents the MgB$_2$ wire produced by Columbus Superconductors used for reference.

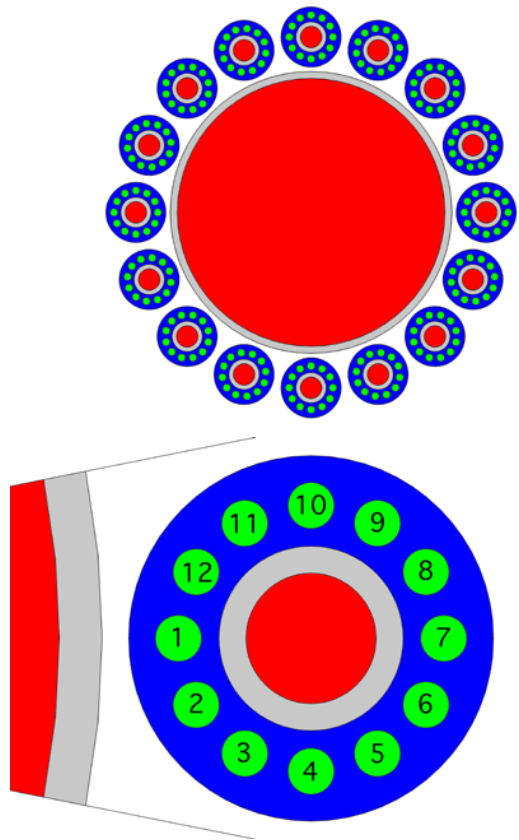

Figure 2. Top: Layout of 1.8 kA power cable consisting of 16 wires, each with 12 filaments of MgB2 (green) in a Monel matrix (blue) and a central copper core (red). The core is surrounded by a Fe layer (grey) to prevent magnetic flux diffusion in copper. Bottom: Modelled 1/16-sector of cross-sectional area containing one strand with the numbering scheme for the filaments used in the text.

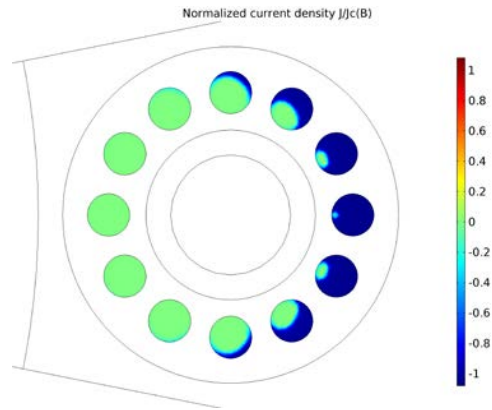

Figure 3. Exemplary current density distribution normalized by the local critical value $J_c(B)$ for the case of straight (coupled) filaments. The applied current in the cable is 2.4 kA (peak). The distribution refers to the peak current.

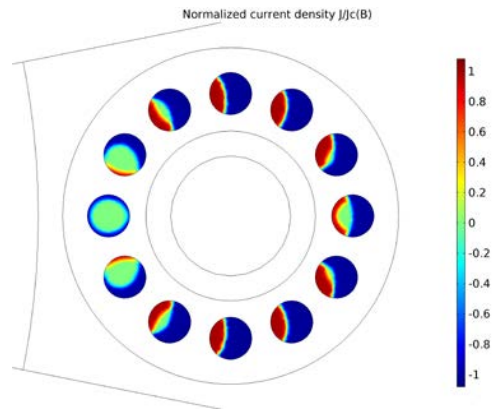

Figure 4. Exemplary current density distribution normalized by the local critical value $J_c(B)$ for the case of twisted (uncoupled) filaments. The applied current in the cable is 2.4 kA (peak). The distribution refers to the peak current.

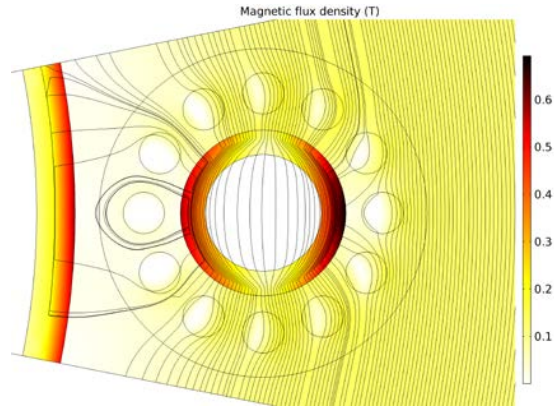

Figure 5. Magnetic flux density distribution for the uncoupled case displayed in Figure 4. The magnetic flux is concentrated in the ferromagnetic parts.

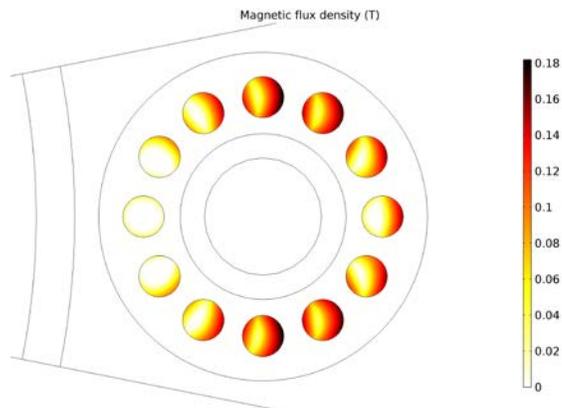

Figure 6. Detailed view of the magnetic flux density distribution in the superconducting filaments only.

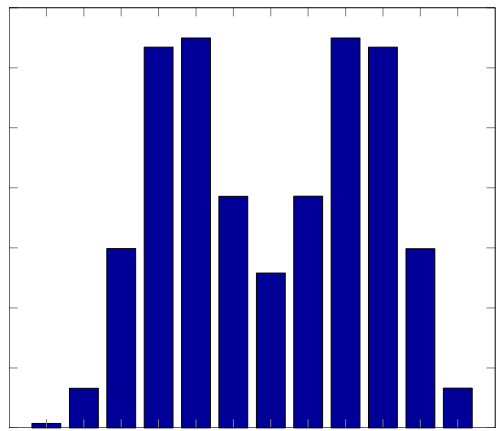

Figure 7. Power loss in the different filament positions (referred to the numbering of Figure 4) for an applied current of 2.4 kA (peak) and $J_{c0}$=3·10$^9$ A/m$^2$. The loss values refer to the whole cable's cross-section.

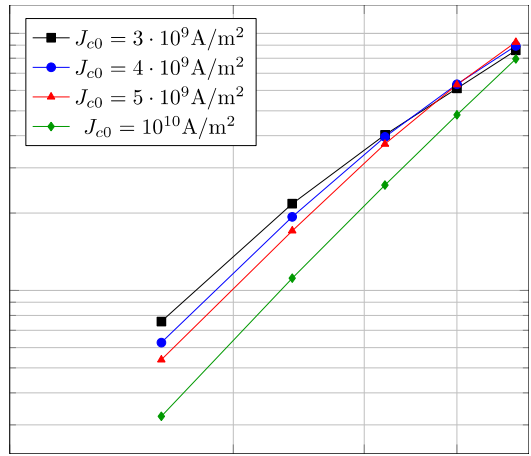

Figure 8. AC power loss at 50 Hz as a function of the transport current for different values of the parameter $J_{c0}$.

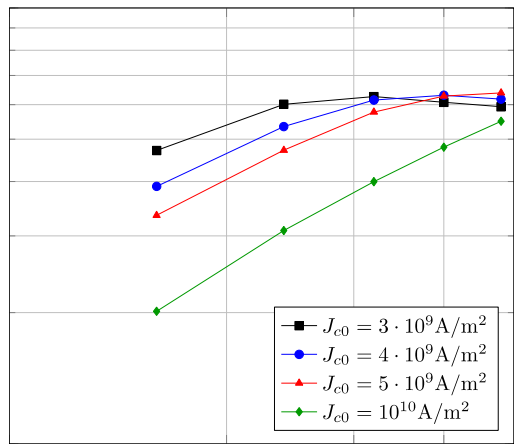

Figure 9. Same AC power loss data of Figure 8, but normalized by the square of the transport current.